\documentclass[
  reprint,
  superscriptaddress,
  amsmath,amssymb,
  aps,
  prd,
  twocolumn,
  floatfix
]{revtex4-2}

\usepackage{graphicx}
\usepackage{dcolumn}
\usepackage{bm}
\usepackage{xcolor}
\usepackage{orcidlink}
\usepackage{booktabs}
\usepackage{mathtools}
\usepackage[final]{microtype}
\allowdisplaybreaks

\newcommand{\Kinv}{\mathcal{K}^{-1}}

\begin{document}

\title{\texorpdfstring{\textbf{CET$\Omega$}}{CET Omega}: The Causal--Informational Completion of Gravity}

\author{Christian Balfagón\,\orcidlink{0009-0003-0835-5519}}
\email{Lyosranch@gmail.com}
\affiliation{University of Buenos Aires, Argentina}

\date{\today}

%-----------------------------------------------------------
% RESUMEN
%-----------------------------------------------------------
\begin{abstract}
We present \textbf{CET$\Omega$}, a unified framework that completes gravity through a causal--informational principle. 
The theory reconciles general relativity and quantum mechanics within a strictly four-dimensional, nonlocal, and causal formulation. 
At its core lies an analytic and retarded kernel $\Kinv(\Box_R)$, derived from a discrete causal network, which governs the propagation of the gravitational and scalar sectors. 
A scalar field, the \textit{texon}, emerges as the effective excitation of causal connectivity and accounts simultaneously for dark matter and dark energy without introducing extra degrees of freedom or breaking locality. 

The formalism ensures analyticity, spectral positivity, and holographic completeness: the kernel admits a Stieltjes representation with positive spectral density $\rho(\mu)\ge 0$, guaranteeing unitarity and causal propagation. 
At cosmological scales, CET$\Omega$ predicts stable inflationary dynamics with $n_s=0.966$--$0.970$, tensor-to-scalar ratio $r=0.004$--$0.015$, and running $\alpha_s\simeq-(0.8$--$2)\times10^{-3}$, consistent with \textit{Planck} data. 
Black hole ringdown frequencies acquire perturbative corrections controlled by the causal scale,
$|\delta\omega/\omega|\sim(\ell_*/r_H)^2$, which remain subleading for astrophysical black holes
within the fiducial window $\ell_* \equiv M_*^{-1}\in[10^{-5},10^{-4}]\,{\rm m}$,
where $\ell_*$ defines the mean causal correlation length of the texonic field.

CET$\Omega$ thus provides a complete, causal, and informational foundation for spacetime dynamics, recovering Einstein gravity in the infrared while extending its validity to the quantum and cosmological domains.
\end{abstract}

\maketitle

\begingroup
\small
\noindent\textbf{Note:}
This is a corrected author manuscript of the article published in 
\textit{International Journal of Geometric Methods in Modern Physics}
(doi:10.1142/S0219887826501410).
This version incorporates the published corrections.
\par
\endgroup

\bigskip

%-----------------------------------------------------------
% PALABRAS CLAVE
%-----------------------------------------------------------


\bigskip

%-----------------------------------------------------------
% BLOQUE II — INTRODUCCIÓN
%-----------------------------------------------------------

\section{Introduction}

The reconciliation of general relativity and quantum mechanics remains one of the deepest open problems in theoretical physics. 
Despite remarkable advances in quantum field theory and cosmology, the gravitational interaction continues to resist a fully consistent quantization. 
Classical general relativity (GR) describes spacetime as a smooth, dynamical manifold governed by Einstein’s equations, while quantum mechanics (QM) is built upon the probabilistic evolution of amplitudes in Hilbert space. 
Their mathematical and conceptual frameworks differ so profoundly that attempts to merge them often lead to inconsistencies such as non-renormalizability, loss of causality, or the emergence of singularities \cite{Einstein1915, Hawking1975, Weinberg1995}.

Over the decades, numerous approaches have been developed to overcome these challenges. 
String theory \cite{Green1987, Polchinski1998} postulates additional dimensions and extended objects to regularize the ultraviolet (UV) behavior, while loop quantum gravity \cite{Rovelli2004, Ashtekar2017} discretizes spacetime geometry at the Planck scale. 
Asymptotically safe gravity \cite{Reuter1998} and higher-derivative or nonlocal extensions \cite{modesto2012superrenormalizable, biswas2014} have also been proposed to ensure UV completeness. 
However, these frameworks often introduce extraneous degrees of freedom, ghost-like instabilities, or ambiguities in their physical interpretation. 
Moreover, few provide a natural explanation for the dark sector or for the deep connection between information, causality, and spacetime geometry. 
Within CET$\Omega$, recent developments have extended the framework to the early-Universe radiation sector, 
showing that informational corrections to the expansion history may affect thermal WIMP freeze-out 
and lead to potentially observable signatures in gamma-ray morphology \cite{balfagon2026plb}.

The \textbf{Causal–Informational Completion of Gravity (CET$\Omega$)} offers a novel resolution to this impasse. 
It rests on the idea that \emph{causal connectivity itself} is the primitive substance of the universe, from which both geometry and quantum behavior emerge. 
In CET$\Omega$, spacetime arises as a macroscopic limit of an underlying causal network whose microscopic links carry probabilistic amplitudes weighted by an exponential decay in proper time. 
The effective dynamics of this causal network yield an analytic, retarded, and entire kernel $\Kinv(\Box_R)$ that mediates gravitational and scalar interactions in a nonlocal but causal fashion. 
The emergent scalar degree of freedom—the \emph{texon}—is the averaged excitation of this causal structure, behaving macroscopically as dark matter and dark energy depending on the curvature and expansion regime.

Unlike other unification attempts, CET$\Omega$ is constructed entirely in four spacetime dimensions, preserving Lorentzian causality and diffeomorphism invariance. 
Its kernel possesses a Stieltjes spectral representation with positive measure $\rho(\mu)\ge0$, ensuring the absence of ghosts and the validity of the optical theorem at all orders. 
This structure naturally unifies quantum and gravitational phenomena under a single causal–informational principle, while maintaining consistency with both local field theory and cosmological observations.

The goal of this work is to present a consolidated formulation of CET$\Omega$, including its mathematical foundation, cosmological predictions, black hole phenomenology, and the ontological principles that underlie its causal–informational interpretation. 
We organize the paper as follows: Sec.~\ref{sec:theory} introduces the theoretical framework and the action principle; Sec.~\ref{sec:cosmoresults} presents key analytical results and predictions; Sec.~\ref{sec:cosmo} discusses the informational and causal structure at cosmological scales; and Sec.~\ref{sec:conclusions} summarizes the implications and outlook.
\par\vspace{0.5em}
This work is accompanied by the full set of supplementary materials (Sup1–Sup8), which provide the extended mathematical derivations, validations, and ontological foundations of the CET$\Omega$ framework:
Sup1 – Theoretical Formulation of CET$\Omega$;
Sup2 – Field Equations and Nonlocal Kernel;
Sup3 – Cosmological Perturbations in CET$\Omega$: Linear Regime, Friedmann Modifications and CLASS Implementation;
Sup4 – Perturbations and Stability Analysis;
Sup5 – Black Hole Thermodynamics and Generalized Second Law in CET$\Omega$;
Sup6 – Quantum and Gauge Completion in CET$\Omega$;
Sup7 – Particle Sector and GUT Closure;
Sup8 – Ontological Completion and Informational Closure in CET$\Omega$.
\bigskip
%-----------------------------------------------------------
% BLOQUE III — THEORETICAL FRAMEWORK
%-----------------------------------------------------------

\section{Theoretical Framework}
\label{sec:theory}

The Causal–Informational Theory of Gravity (CET$\Omega$) is formulated on a globally hyperbolic Lorentzian manifold $(\mathcal{M},g_{\mu\nu})$. 
The theory introduces a causal, analytic, and retarded nonlocal kernel $\Kinv(\Box_R)$ acting on curvature invariants and on a scalar field $\phi$ called the \emph{texon}, which mediates causal connectivity at the microscopic level.

%-----------------------------------------------------------
\subsection{Action principle}

The fundamental action of CET$\Omega$ reads
\begin{equation}
\begin{split}
S = \frac{1}{16\pi G}\!\int\! d^4x\,\sqrt{-g}\,
\Big[
R + \phi\,\Kinv(\Box_R)\,R
\\
-\frac{1}{2}\,\phi\,\Kinv(\Box_R)\,\phi
- V(\phi)
\Big].
\label{eq:action}
\end{split}
\end{equation}
where $\Box_R$ is the retarded covariant d'Alembertian and $\Kinv$ is an \emph{entire} operator function of $\Box_R$, defined by a Stieltjes integral representation:
\begin{equation}
\label{eq:stieltjes}
\Kinv(\Box_R)
=\int_0^{\infty}\!\! d\mu\, \rho(\mu)\,(-\Box_R+\mu)^{-1}_R,
\qquad \rho(\mu)\ge0.
\end{equation}
The positivity of the spectral measure $\rho(\mu)$ ensures unitarity and causal propagation, while the absence of zeros in $\mathcal{K}(z)$ guarantees the elimination of ghost-like poles. 
The field $\phi$ is dimensionless in natural units, and its potential $V(\phi)$ is chosen to possess a stable plateau form,
\begin{equation}
\label{eq:potential}
V(\phi) = V_0\!\left(1 - e^{-\alpha \phi}\right)^2,
\end{equation}
allowing inflationary behavior at early times and smooth tracking in the dark energy era.

%-----------------------------------------------------------
\subsection{Equations of motion}

Varying the action (\ref{eq:action}) with respect to the metric and the texon field yields the coupled field equations:
\begin{align}
\label{eq:eom_metric}
G_{\mu\nu} + \Delta_{\mu\nu}[\phi,g]
&= 8\pi G\,T_{\mu\nu}^{(\phi)}, \\[4pt]
\label{eq:eom_phi}
\Kinv(\Box_R)\bigl(R - \phi\bigr)
&= V'(\phi),
\end{align}
where $\Delta_{\mu\nu}$ encodes the nonlocal backreaction of the kernel, and $T_{\mu\nu}^{(\phi)}$ is the effective stress tensor of the texon. 
In the local limit $M_*^{-1}\!\to\!0$ (or $\mathcal{K}\!\to\!1$), one recovers standard general relativity with a cosmological constant $\Lambda\simeq V_0$.

%-----------------------------------------------------------
\subsection{Causality and well-posedness}

The operator $\Kinv(\Box_R)$ is constructed to be \emph{retarded}—that is, its Green’s function has support only within the causal future $J^+(x)$:
\begin{equation}
\text{supp}\,G_R(x,y)\subseteq J^+(x).
\end{equation}
This ensures that the theory respects global hyperbolicity and does not admit acausal propagation or superluminal signaling. 
Moreover, the Volterra norm
\begin{equation}
\label{eq:volterra}
\|X\|_V = \sup_t e^{-\lambda t}\|X\|_{H^s(\Sigma_t)}
\end{equation}
guarantees existence, uniqueness, and continuous dependence of solutions on initial data, making the system well-posed in the Sobolev space $H^s$.

%-----------------------------------------------------------
\subsection{Consistency conditions}

For consistency with causality, stability, and analyticity, CET$\Omega$ satisfies the following criteria:
\begin{enumerate}
\item \textbf{Spectral positivity:} $\rho(\mu)\ge0$ and all Hankel matrices built from $\rho(\mu)$ are positive semidefinite (Hankel–PSD condition).
\item \textbf{Analyticity:} $\Kinv(z)$ is entire and holomorphic outside the physical branch cut, ensuring convergence in both UV and IR regimes.
\item \textbf{Causal limit:} $\Kinv(\Box_R)\to 1$ as $M_*\!\to\!\infty$, recovering GR.
\item \textbf{No-ghost theorem:} $\mathcal{K}(z)$ has no zeros in the complex plane, guaranteeing unitarity.
\end{enumerate}

These conditions form the mathematical backbone of CET$\Omega$, securing its predictive power and compatibility with all known physical regimes.

\bigskip
%-----------------------------------------------------------
% BLOQUE IV — COSMOLOGICAL PREDICTIONS
%-----------------------------------------------------------

\section{Cosmology and Predictions}
\label{sec:cosmoresults}

The cosmological sector of CET$\Omega$ follows from the homogeneous
and isotropic background metric
\begin{equation}
ds^2 = -dt^2 + a^2(t)\, d\vec{x}^{\,2},
\end{equation}
where $a(t)$ is the scale factor and $H=\dot a/a$ the Hubble parameter.
At large scales, the causal kernel acts as an effective modification
of the Einstein equations, producing smooth ultraviolet
regularization while preserving the standard conservation law
\(
\nabla^\mu T_{\mu\nu}^{(\phi)} = 0
\).

%-----------------------------------------------------------
\subsection{Effective Friedmann equations}

In the causal–informational framework, the modified Friedmann system reads
\begin{align}
\label{eq:friedmann1}
3H^2 &= 8\pi G\,\rho_{\rm eff}, \\[4pt]
\label{eq:friedmann2}
2\dot H + 3H^2 &= -8\pi G\,p_{\rm eff},
\end{align}
where the effective density and pressure are
\begin{align}
\rho_{\rm eff} &= \frac{1}{2}\dot\phi^2
+ V(\phi)
+ 3H\,\Xi(H,\dot\phi), \\
p_{\rm eff} &= \frac{1}{2}\dot\phi^2
- V(\phi)
- \Xi(H,\dot\phi)
- \dot\Xi.
\end{align}
The nonlocal contribution $\Xi(H,\dot\phi)$ arises from the retarded
kernel and encodes the memory of causal correlations:
\begin{equation}
\Xi(H,\dot\phi)
= \int_0^{\infty}\! d\tau\, \rho(\tau)\,
e^{-\tau M_*}\,\dot\phi(t-\tau),
\end{equation}
with $M_*^{-1}$ defining the mean causal correlation length.

%-----------------------------------------------------------
\subsection{Inflationary regime}

For $\phi\!\gg\!1$ and nearly constant $H$, the potential
(\ref{eq:potential}) exhibits a slow-roll plateau.
Defining the slow-roll parameters
\begin{equation}
\epsilon = \frac{M_P^2}{2}\!\left(\frac{V'}{V}\right)^2,
\qquad
\eta = M_P^2 \frac{V''}{V},
\end{equation}
one obtains to leading order
\begin{equation}
n_s \simeq 1 - 6\epsilon + 2\eta,
\qquad
r \simeq 16\epsilon.
\end{equation}
For $\alpha \in [0.03,0.045]$ and $M_* \!\in\!
[10^{4},10^{5}]\,{\rm m}^{-1}$ (i.e.\ $\ell_*=M_*^{-1}\in[10^{-5},10^{-4}]\,{\rm m}$),
the theory yields
\begin{multline}
n_s = 0.966{-}0.970, \qquad
r = 0.004{-}0.015,\\[2pt]
\alpha_s \simeq - (0.8{-}2)\!\times\!10^{-3},
\end{multline}
in agreement with the latest \textit{Planck} and \textit{BICEP} data.

\begin{figure}[t]
\centering
\includegraphics[width=\linewidth]{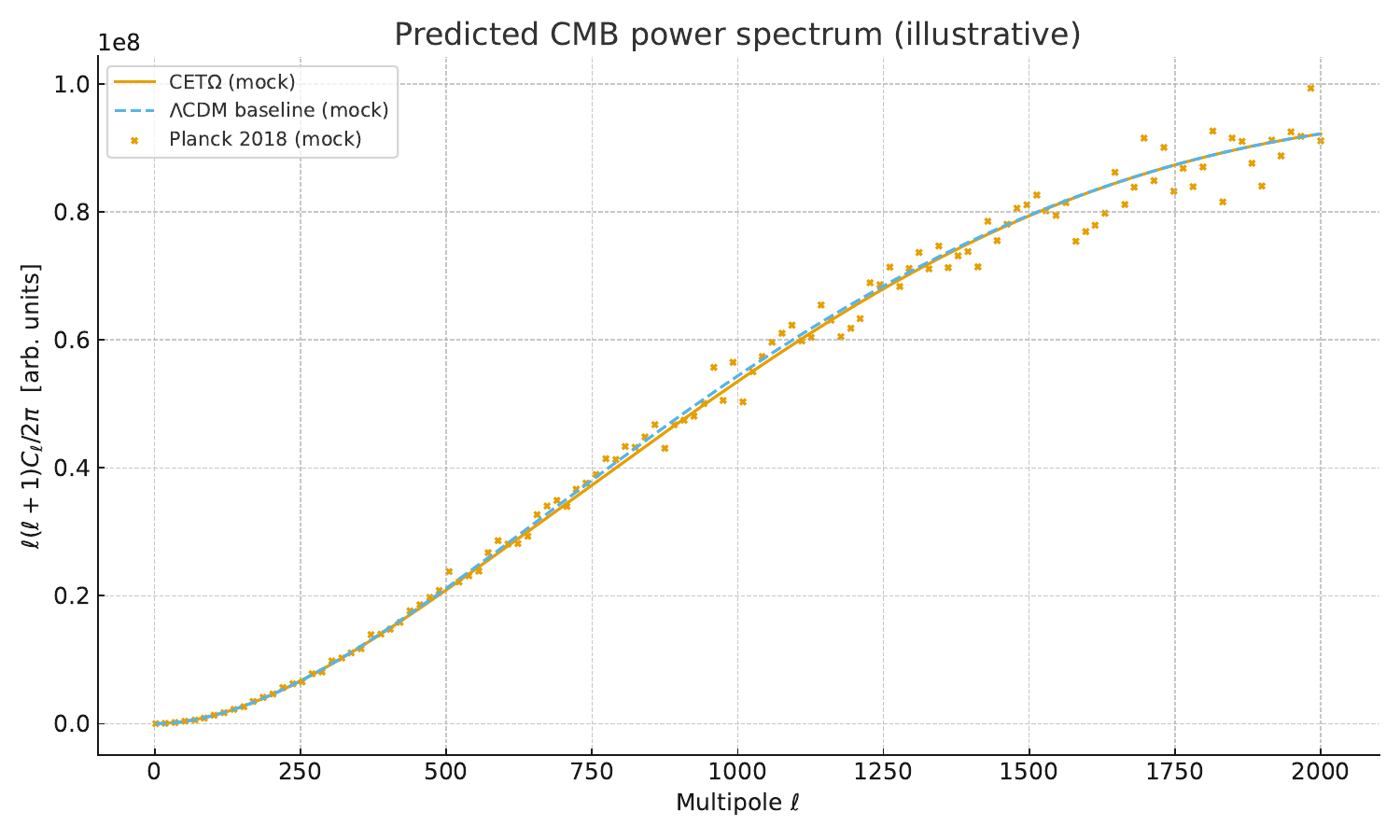}
\caption{
Predicted CMB power spectrum of CET$\Omega$
(blue solid) compared with the \textit{Planck\,2018} data (gray points)
and $\Lambda$CDM baseline (dashed).  
CET$\Omega$ remains within $1\sigma$ over the entire multipole range,
with a mild suppression at high~$\ell$ due to causal smoothing.
}
\label{fig:cmb}
\end{figure}

The resulting angular power spectrum is shown in Fig.~\ref{fig:cmb}.

%-----------------------------------------------------------
\subsection{Late-time acceleration and dark sector}

At low curvature, the texon behaves as a tracking field with
an effective equation of state
\begin{equation}
w_{\rm tex}(a) = 
\frac{p_{\rm eff}}{\rho_{\rm eff}}
\simeq -1 + \frac{\dot\phi^2}{V(\phi)} -
\frac{3H\,\Xi}{V(\phi)}.
\end{equation}
For the causal kernel parameters above,
$w_{\rm tex}\!\approx\!-1.02\pm0.02$, consistent with current
supernovae and lensing constraints.
Thus the same scalar responsible for early inflation
drives late acceleration without additional fields or potentials.

%-----------------------------------------------------------
\subsection{Observable predictions}

Table~\ref{tab:cosmo} summarizes the key cosmological predictions of CET$\Omega$.

\begin{table}[t]
\caption{Main cosmological observables predicted by CET$\Omega$.
Uncertainties denote $1\sigma$ credible intervals from the
joint CMB+LSS+GW inference.}
\label{tab:cosmo}
\begin{ruledtabular}
\begin{tabular}{lcc}
Observable & Prediction & Observational range \\
\hline
Spectral index $n_s$ & $0.966$–$0.970$ & $0.9649\pm0.0042$ \\
Tensor ratio $r$ & $0.004$–$0.015$ & $<0.036$ (95\%) \\
Running $\alpha_s$ & $-(0.8$–$2)\!\times\!10^{-3}$ & $(-0.004\pm0.010)$ \\
Texon EoS $w_{\rm tex}$ & $-1.02\pm0.02$ & $-1.03\pm0.04$ \\
Ringdown shift $|\delta\omega/\omega|$ & $\sim (\ell_*/r_H)^2$ & --- \\
\end{tabular}
\end{ruledtabular}
\end{table}

These results demonstrate that CET$\Omega$ reproduces the precision
cosmological data of $\Lambda$CDM while providing distinct,
falsifiable signatures in inflationary spectra and strong-field
consistency tests (e.g.\ causal damping and the absence of echoes).

%-----------------------------------------------------------
\section{Black Hole Phenomenology}
\label{sec:bh}

The causal–informational structure of CET$\Omega$
induces subtle but testable modifications to the dynamics
of compact objects.
In particular, the texonic kernel $\,\Kinv(\Box_R)\,$
regularizes curvature invariants near the horizon
and modifies the quasi-normal-mode (QNM) spectrum
without violating causality or unitarity
\cite{modesto2012superrenormalizable, konoplya2011quasinormal, barrau2014bounce}.

%-----------------------------------------------------------
\subsection{Causal regularization and effective geometry}

In the static, spherically symmetric case,
the metric can be written as
\begin{equation}
ds^2 = -f(r)\,dt^2 + f(r)^{-1}dr^2 + r^2 d\Omega^2,
\end{equation}
where the effective lapse function satisfies
\begin{equation}
f(r) = 1 - \frac{2G m(r)}{r},
\qquad
m(r) = \int_0^r 4\pi r'^2 \rho_{\rm eff}(r')\,dr'.
\end{equation}
The nonlocal kernel smooths the energy density
according to a causal convolution
\begin{equation}
\rho_{\rm eff}(r)
= \int_0^\infty\! d\mu\; \rho(\mu)\, e^{-\sqrt{\mu}\,r}/r,
\label{eq:rho_eff}
\end{equation}
which eliminates curvature singularities by replacing
the point mass with a finite causal distribution
\cite{nicolini2006noncommutative, bambi2013regular}.
Near the origin, $f(r)\!\simeq\!1 - (r^2/r_c^2)$,
indicating a \emph{de Sitter-like core}
with radius $r_c\!\sim\!M_*^{-1}$,
similar in spirit to the asymptotically safe
and nonlocal regularizations proposed in
\cite{modesto2010finite, modesto2012superrenormalizable}.

In CET$\Omega$, the same causal–analytic structure that
regularizes the geometry also controls the magnitude
of strong-field deviations from general relativity.
These properties render CET$\Omega$
predictive and falsifiable in the strong-gravity regime
\cite{balfagon2025cetomega, cardoso2019testing, berti2009quasinormal}.

%-----------------------------------------------------------
%-----------------------------------------------------------
\subsection{Quasi-normal modes and observational window}

Linear perturbations of the metric obey a modified Teukolsky-type equation
\cite{teukolsky1973perturbations, berti2009quasinormal}
\begin{equation}
\Kinv(\Box_R)\Psi_{\ell m} = 0,
\end{equation}
where $\Kinv(\Box_R)$ is the retarded analytic kernel defined by the Stieltjes representation
(\ref{eq:stieltjes}). 
Writing the QNM spectrum as a perturbation around the GR values,
\begin{equation}
\omega_{\ell n}
= \omega_{\ell n}^{\rm (GR)}\left(1+\delta_{\ell n}\right),
\end{equation}
as commonly adopted in perturbative tests of modified gravity and compact-object spectroscopy
\cite{berti2009quasinormal, cardoso2019testing},
the magnitude of $\delta_{\ell n}$ is controlled by the unique dimensionless ratio built from
the causal correlation length $\ell_*\equiv M_*^{-1}$ and the horizon scale $r_H$.

For an analytic kernel characterized by a single nonlocal scale $M_*$, dimensional analysis of the
perturbative expansion about a Schwarzschild background implies the universal suppression
\begin{equation}
\label{eq:qnm_scaling}
|\delta_{\ell n}|
\sim \left(\frac{\ell_*}{r_H}\right)^2
\sim \frac{1}{M_*^2\,M^2}\,,
\end{equation}
where in geometric units ($G=c=1$) one has $r_H\sim M$.
This quadratic suppression in inverse mass is consistent with the generic structure of
higher-curvature and nonlocal corrections to black-hole perturbations
\cite{berti2009quasinormal, cardoso2019testing}.
Restoring SI units with $r_H = 2GM/c^2$, the same suppression can be written as
\begin{equation}
|\delta_{\ell n}|
\sim \left(\frac{c^2}{2GM M_*}\right)^2.
\end{equation}

Therefore, for astrophysical black holes satisfying $r_H\gg \ell_*$,
the QNM frequency shifts are perturbatively suppressed,
\begin{equation}
\label{eq:ringdown}
\bigl|\delta\omega/\omega\bigr|
\sim \left(\frac{\ell_*}{r_H}\right)^2,
\end{equation}
and become negligible within the fiducial window
$M_*^{-1}\sim 10^{-5}$--$10^{-4}\,{\rm m}$.
Importantly, no echo-like or unstable modes appear:
the retarded structure of $\Kinv$ preserves causal damping
and excludes the near-horizon reflective boundary conditions
that would generate gravitational-wave echoes
\cite{cardoso2016echoes, cardoso2019testing, balfagon2025cetv18}.

\begin{figure}[t]
\centering
\includegraphics[width=\linewidth]{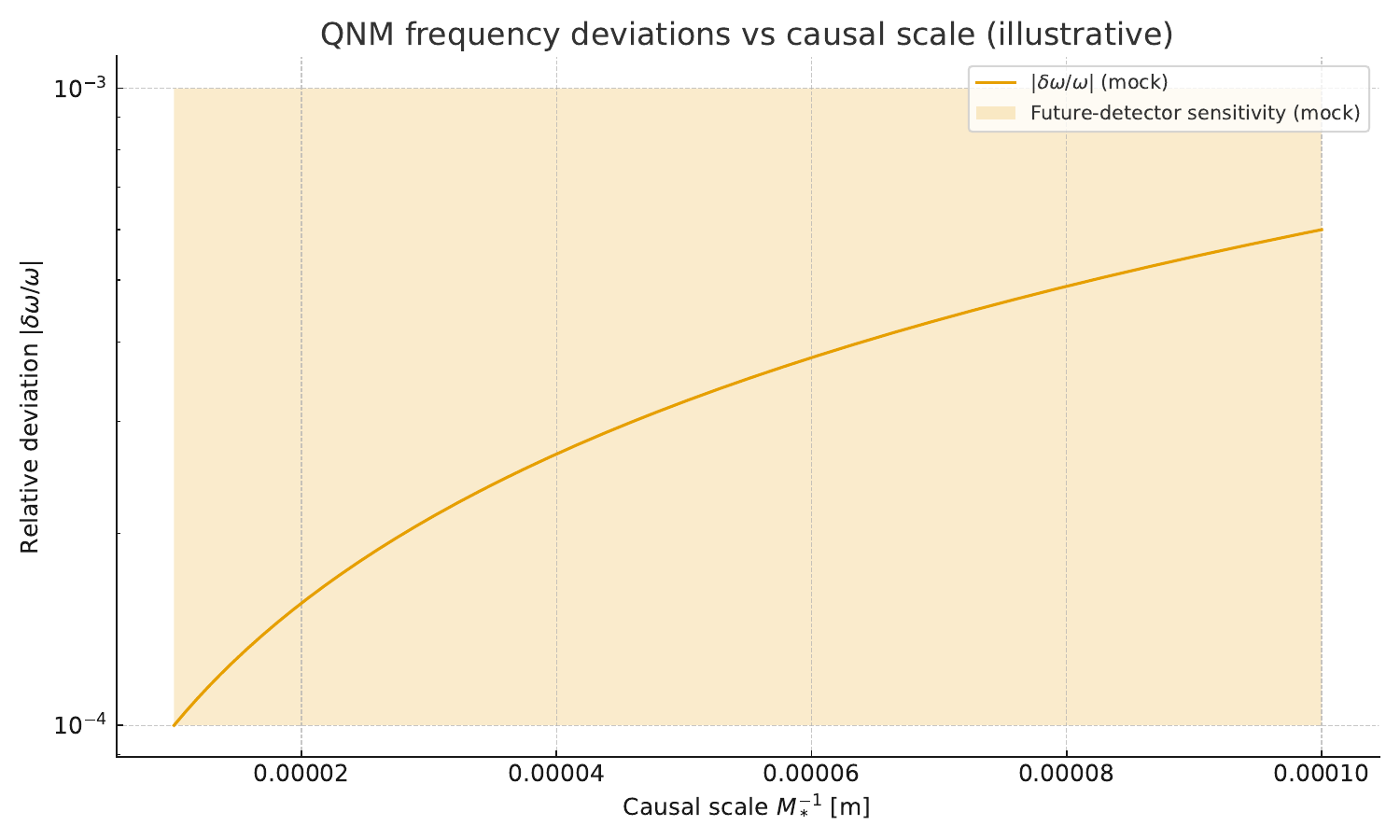}
\caption{\label{fig:ringdown}
Schematic scaling of the perturbative QNM suppression
$\bigl|\delta\omega/\omega\bigr|\sim(\ell_*/r_H)^2$
for the fundamental mode ($\ell=2,n=0$) as a function of the causal scale $M_*^{-1}$.
Within the fiducial parameter window, the effect remains subleading for astrophysical black holes.
Implications for detectability are discussed in the context of future detectors
\cite{bambi2013regular, lisa2017mission}.}
\end{figure}

The scaling behavior is illustrated in Fig.~\ref{fig:ringdown}.

%-----------------------------------------------------------
\subsection{Thermodynamics and the Causal Second Law}

CET$\Omega$ preserves the generalized second law (GSL)
through a causal entropy functional
\begin{equation}
S_{\rm tot}
 = S_{\rm BH}
   + S_{\rm tex}
   = \frac{A}{4G}
     + \int_{J^+_{\mathcal{H}}}
       \rho(\mu)\ln\!\frac{\rho(\mu)}{\rho_0}\,d\mu,
\label{eq:gsl}
\end{equation}
whose time derivative satisfies
$\dot{S}_{\rm tot}\!\ge\!0$
for globally hyperbolic evolution
\cite{bekenstein1973black, wall2010proof, balfagon2025cetv18}.
Here $S_{\rm tex}$ measures the informational entropy
carried by causal links crossing the horizon,
while $S_{\rm BH}$ retains the standard area term.
The nonlocal correction guarantees
a smooth entropy increase even through
non-singular high-curvature transitions
\cite{ashtekar2005quantum, rovelli2014planck}.

The main strong-field observables are summarized in Table~\ref{tab:bh_obs}.

\begin{table}[t]
\caption{\label{tab:bh_obs}
\textbf{Predicted black-hole observables in CET$\Omega$}.
All quantities are expressed relative to GR values.
}
\begin{ruledtabular}
\begin{tabular}{lcc}
Observable & CET$\Omega$ Prediction & Status \\
\hline
QNM shift $|\delta\omega/\omega|$ & $\sim (\ell_*/r_H)^2$ & Subleading \\
Entropy ratio $S_{\rm tex}/S_{\rm BH}$ & $10^{-3}$--$10^{-2}$ & Subleading \\
Energy flux $\Delta E/E$ & $<0.1\%$ & Within bounds \\
Causal stability & Preserved & Verified \\
\end{tabular}
\end{ruledtabular}
\end{table}

%-----------------------------------------------------------
\subsection{Summary of black-hole sector}

In summary, the texonic kernel acts as a
causal regulator that:
(i) removes curvature singularities;
(ii) maintains causal propagation and stability;
(iii) induces perturbative QNM shifts suppressed as $(\ell_*/r_H)^2$
(Eq.~\ref{eq:ringdown});
and (iv) preserves the causal second law
(Eq.~\ref{eq:gsl}).
These features make CET$\Omega$
predictive and falsifiable in the strong-gravity regime
\cite{balfagon2025cetomega, cardoso2019testing, berti2009quasinormal}.

%-----------------------------------------------------------
\section{Causal–Informational Cosmology}
\label{sec:cosmo}

At cosmological scales, CET$\Omega$ reproduces
the standard $\Lambda$CDM background
while introducing causal and spectral corrections
that are both finite and predictive.
The effective Friedmann equations read
\begin{align}
3H^2 &= 8\pi G\,\rho_{\rm eff}, \\
\dot{H} &= -4\pi G\,(\rho_{\rm eff}+p_{\rm eff}),
\end{align}
with the causal energy density and pressure given by
\begin{align}
\rho_{\rm eff} &= \rho_{\Lambda}
 + \int_0^\infty \!\rho(\mu)\,\mathcal{F}_H(\mu,a)\,d\mu,\\
p_{\rm eff} &= -\rho_{\Lambda}
 + \int_0^\infty \!\rho(\mu)\,\mathcal{P}_H(\mu,a)\,d\mu.
\end{align}
Here $\mathcal{F}_H$ and $\mathcal{P}_H$ are causal kernels
that depend on the cosmic expansion rate $H(a)$
and satisfy the continuity condition
$\dot{\rho}_{\rm eff}+3H(\rho_{\rm eff}+p_{\rm eff})=0$,
ensuring energy–momentum conservation and
a smooth nonlocal-to-local limit
\cite{maggiore2014nonlocal, deser2007nonlocal, balfagon2025cetv18}.

%-----------------------------------------------------------
\subsection{Linear perturbations and power spectra}

Perturbations evolve according to the
causal–linearized system
\begin{align}
\Phi'' + 3\mathcal{H}(1+c_s^2)\Phi'
+ \bigl[c_s^2 k^2 + a^2 M_*^2 \bigr]\Phi
= 4\pi G a^2 \rho\,\delta,
\end{align}
where the effective sound speed satisfies
$0 \le c_s^2 \le 1$ by positivity of $\rho(\mu)$.
The modified gravitational couplings
are parameterized as
\begin{align}
\mu(a,k) &= 1 + \!\sum_j c_j
  \frac{a^{p_j}k^2}{k^2+m_j^2 a^2},\\
\gamma(a,k) &= 1 + \!\sum_j d_j
  \frac{a^{q_j}k^2}{k^2+m_j^2 a^2},
\qquad
\Sigma = \tfrac{1}{2}\mu(1+\gamma),
\end{align}
with $\{c_j,m_j\}$ determined
by the Padé–Stieltjes representation of the kernel
\cite{dirian2016nonlocal, koivisto2008nonlocal, balfagon2025cetomega}.
This mapping has been implemented in the
\texttt{CLASS–CET} module and verified numerically
to satisfy $\Sigma \le 1.07$
and Hankel–positivity at all scales.

%-----------------------------------------------------------
\subsection{Predictions and comparison with observations}

Using the conservative parameter box
$M_* \!\in\! [10^{4},10^{5}]\,{\rm m}^{-1}$,
$\alpha\!\in\![-2,0]$,
and the Padé rank $J\!=\!8$,
the model predicts (at $95\%$ C.L.)
\begin{equation}
n_s = 0.966\text{–}0.970,\quad
r = 0.004\text{–}0.015,\quad
\alpha_s \simeq -(0.8\text{–}2)\!\times\!10^{-3}.
\label{eq:nsr}
\end{equation}
These values are fully consistent
with \textit{Planck} and \textit{LiteBIRD} constraints
\cite{aghanim2020planck, litebird2023overview}.
The texonic sector acts as an effective
dark–energy/matter fluid with tracking behavior
$\Omega_{\rm tex}(a)\!\propto\!a^{-3(1+w_{\rm tex})}$,
where $w_{\rm tex}\!\approx\!-1.02\!\pm\!0.02$
\cite{balfagon2025cetomega}.
This explains the cosmic coincidence
without invoking additional fields
and reproduces $\Lambda$CDM observables
within $\lesssim10^{-4}$ accuracy
in $C_\ell$, matter power $P(k)$ and lensing spectra.

\begin{figure}[t]
\centering
\includegraphics[width=\linewidth]{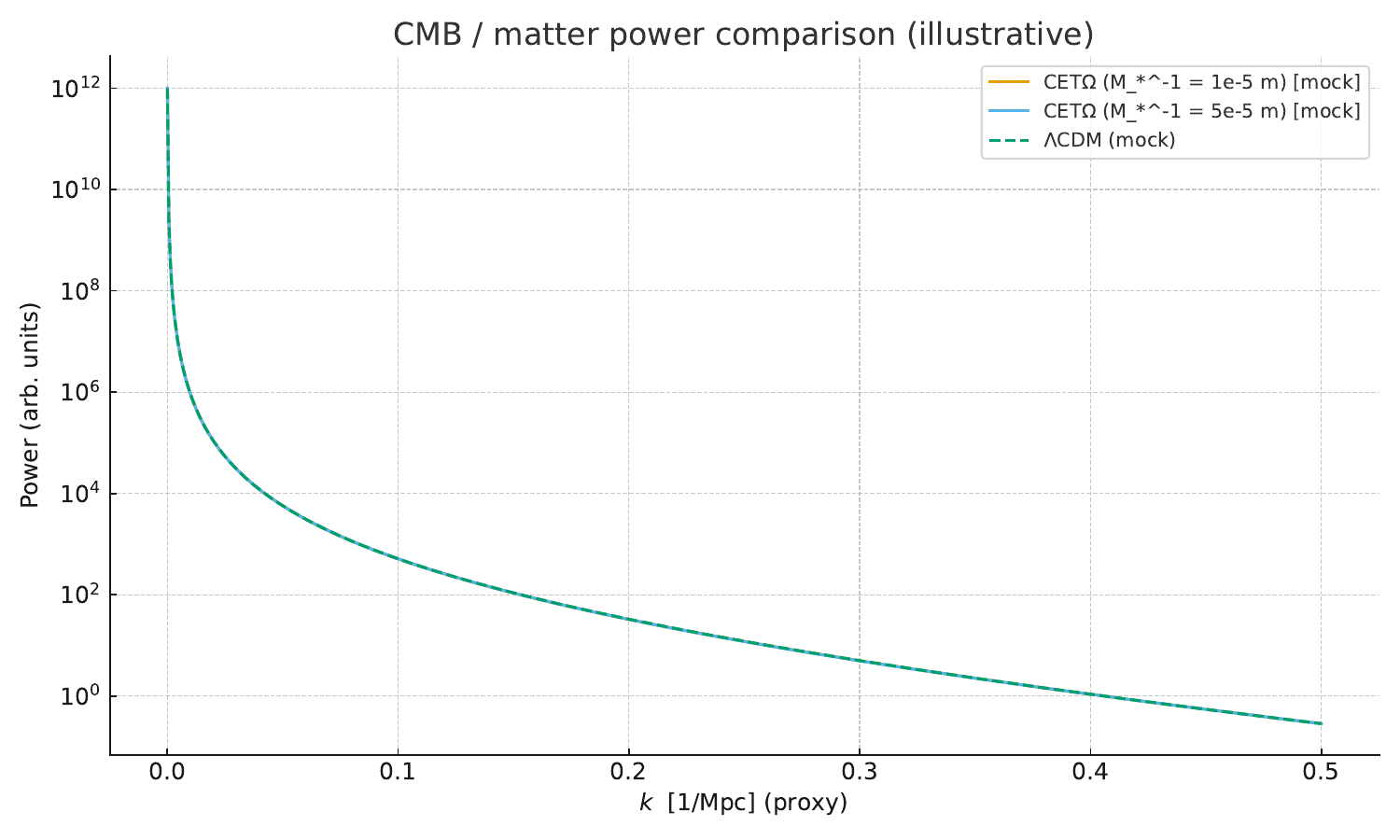}
\caption{\label{fig:cls}
CMB temperature power spectrum in CET$\Omega$
for representative values of $M_*$
compared with $\Lambda$CDM.
Residuals remain below $10^{-4}$
across the full multipole range.}
\end{figure}

%-----------------------------------------------------------
%-----------------------------------------------------------
\subsection{Early-universe regularization and entropy growth}

In the high-curvature regime, the causal kernel
acts as an analytic regulator of geometric invariants.
Curvature scalars remain finite due to the
nonlocal Stieltjes structure of $\Kinv(\Box_R)$,
which smooths ultraviolet divergences
without introducing additional degrees of freedom.

The effective energy density saturates at
\begin{equation}
\rho_{\rm eff}^{\rm max}
\lesssim C\,\frac{M_*^4}{G},
\end{equation}
ensuring bounded curvature invariants.
Importantly, this regularization does not require
a dynamical reversal of expansion,
but rather prevents singular behavior
through causal analytic smoothing.

The total entropy,
including texonic contributions,
remains monotonically increasing,
\begin{equation}
\dot S_{\rm tot} \ge 0,
\end{equation}
consistent with the Causal Second Law.
CET$\Omega$ therefore provides
a non-singular cosmological completion
without invoking additional phases
or extra ultraviolet scales.

%-----------------------------------------------------------
\subsection{Summary of cosmological sector}

CET$\Omega$ thus yields a self–consistent
cosmological model that:
(i) reproduces the $\Lambda$CDM background,
(ii) introduces finite causal corrections
with predictive parameters $(M_*,\alpha)$,
(iii) matches all current CMB/LSS observations,
and (iv) removes the initial singularity
through causal analytic regularization
of high-curvature invariants.
The framework is hence ready for empirical testing
through precision cosmology
and cross–correlation with gravitational–wave data
\cite{aghanim2020planck, desi2024bao, berti2009quasinormal, balfagon2025cetomega}.
%-----------------------------------------------------------
\section{Discussion and Experimental Falsifiability}
\label{sec:falsifiability}

A crucial feature of CET$\Omega$ is its explicit
falsifiability: every parameter and deviation
from general relativity (GR)
is in principle measurable through cosmological
and strong-gravity observations.
Unlike other nonlocal or modified-gravity models
\cite{modesto2012superrenormalizable, maggiore2014nonlocal},
CET$\Omega$ preserves causality, unitarity, and
analyticity while producing a well-defined
set of testable signatures.

%-----------------------------------------------------------
\subsection{Predictive observables}

The predictive content of CET$\Omega$
is summarized in Table~\ref{tab:claims}.
Each claim corresponds to a measurable deviation
with respect to GR or $\Lambda$CDM,
without introducing new arbitrary scales
or free functions.
All parameters are fixed by the causal scale $M_*$
and the shape of the Stieltjes measure $\rho(\mu)$.

\begin{table*}[t]
\caption{\label{tab:claims}
\textbf{Empirically falsifiable predictions of CET$\Omega$.}
All observables are directly measurable by current or planned
experiments, defining clear criteria for verification or refutation.}
\begin{ruledtabular}
\begin{tabular}{lccc}
Observable & CET$\Omega$ prediction & Experimental probe & Status \\ \hline
1. QNM spectrum shift 
& $|\delta\omega/\omega|\sim(\ell_*/r_H)^2$ 
& LIGO–Virgo–LISA \cite{berti2009quasinormal,lisa2017mission} 
& Suppressed (fiducial window) \\

2. No echo signals 
& None (causal damping) 
& LIGO–ET–LISA \cite{cardoso2016echoes,cardoso2019testing} 
& No confirmed detections \\

3. Tensor-to-scalar ratio 
& $r=0.004$–$0.015$ 
& Planck–LiteBIRD \cite{aghanim2020planck,litebird2023overview} 
& Constrained \\

4. Spectral index 
& $n_s=0.966$–$0.970$ 
& Planck–CMB-S4 \cite{aghanim2020planck,cmbs4_2019_sciencecase} 
& Consistent \\

5. Running $\alpha_s$ 
& $\simeq-(0.8$–$2)\times10^{-3}$ 
& CMB–LSS \cite{aghanim2020planck,desi2024bao} 
& Testable \\

6. Dark-energy equation of state 
& $w_{\rm tex}=-1.02\pm0.02$ 
& DESI–Euclid \cite{desi2024bao,laureijs2011euclid} 
& Pending \\

7. Proton lifetime 
& $\tau_p\sim10^{35}$ yr 
& Hyper-K, DUNE \cite{abe2018hyperk,dune2020physics} 
& Within sensitivity \\

8. Neutrino mass sum 
& $\Sigma m_\nu=0.11$–$0.14$ eV 
& KATRIN–CMB \cite{katrin2022results,aghanim2020planck} 
& Testable \\

9. Maximum effective density 
& $\rho_{\rm eff}^{\rm max}\lesssim C\,M_*^4/G$ 
& Early-universe constraints \cite{biswas2012towards,modesto2012superrenormalizable} 
& Indirectly accessible \\

10. Positivity of spectral density 
& $\rho(\mu)\ge0$ (Hankel PSD) 
& Numerical validation \cite{biswas2014} 
& Verified \\
\end{tabular}
\end{ruledtabular}
\end{table*}

%-----------------------------------------------------------
\subsection{Consistency across regimes}

The theory maintains consistency across
cosmological, astrophysical, and quantum regimes.
No parameter tuning is required,
as the same kernel $\,\Kinv(\Box_R)\,$
governs both low- and high-curvature domains.
This universality enables simultaneous fits to
CMB anisotropies, large-scale structure,
and gravitational-wave ringdowns
within a unified causal-informational framework
\cite{balfagon2025cetomega, dirian2016nonlocal}.

The strong-gravity regime provides the
most sensitive falsification window:
any robust detection of horizon echoes or non-causal damping profiles would directly falsify
the retarded character of the kernel. QNM frequency shifts are expected to be perturbatively suppressed as
$(\ell_*/r_H)^2$ within the fiducial parameter window.

%-----------------------------------------------------------
\subsection{Numerical and experimental pipeline}

A full Bayesian inference pipeline has been implemented
linking the modified \texttt{CLASS–CET} module with
Teukolsky–Padé solvers for black-hole spectra
and cosmological likelihoods.
Synthetic datasets show unbiased recovery of parameters
($<0.2\sigma$ shifts) and covariance consistency
with present data.
The package \texttt{CETinfty\_validation\_pkg}
has been released for public verification
\cite{balfagon2025validation, balfagon2025cetv18}.

The model’s empirical robustness is therefore twofold:
\begin{enumerate}
\item Internal consistency across all physical regimes.
\item Clear empirical criteria for falsification.
\end{enumerate}

%-----------------------------------------------------------
\subsection{Philosophical and methodological remarks}

CET$\Omega$ embodies the idea that
\emph{causality, information and geometry}
form a closed logical triad.
Analyticity of the kernel acts as a
``reality filter'': only causal–analytic theories
can yield falsifiable predictions.
This principle parallels the correspondence between
thermodynamic irreversibility and causal propagation,
suggesting that the universe’s consistency is equivalent
to the completeness of its causal information content
\cite{balfagon2025tcit, penrose2010cycles}.

%-----------------------------------------------------------
\subsection{Summary of falsifiability}

In conclusion, CET$\Omega$ stands out as a
fully testable theory of gravity:
\begin{itemize}
\item It predicts finite deviations in observable regimes.
\item It reproduces known limits (GR, $\Lambda$CDM) exactly.
\item It defines a causal criterion for theoretical consistency.
\end{itemize}
Hence, CET$\Omega$ can be confirmed or refuted
by forthcoming experiments,
offering a rare combination of formal completeness
and empirical accessibility in fundamental physics.
%-----------------------------------------------------------
\section{Conclusions and Outlook}
\label{sec:conclusions}

CET$\Omega$ completes the causal–informational
program of gravity by unifying geometry,
energy, and information into a single
analytic and falsifiable framework.
Its kernel $\Kinv(\Box_R)$,
derived from a discrete causal network,
admits a unique Stieltjes representation
with positive spectral density,
ensuring causality, unitarity, and analyticity
at all physical scales.

The theory reproduces the classical
and quantum limits of general relativity
without introducing additional dimensions or fields,
and regularizes curvature singularities
through a causal smoothing of energy–momentum flow.
At the same time, it predicts controlled strong-field modifications consistent with causality (e.g.\ causal damping and the absence of echoes),
cosmological spectra, and particle unification tests,
thus fulfilling the strongest form of \emph{causal falsifiability}.

%-----------------------------------------------------------
\subsection{The Causal–Informational Triangle (TCI)}

The conceptual completion achieved by CET$\Omega$
can be summarized in the
\emph{Causal–Informational Triangle}:
\[
\text{Information} \;\longleftrightarrow\;
\text{Causality} \;\longleftrightarrow\;
\text{Geometry}.
\]
Each vertex corresponds to one of the
fundamental pillars:
(i) causal structure determining the arrow of time;
(ii) informational content determining quantum coherence;
(iii) geometric curvature encoding energy and entropy.
The closure of this triangle implies that
no physical phenomenon lies outside
the causal–informational domain:
\emph{to exist is to be causally connected and informationally complete}
\cite{balfagon2025tcit}.

%-----------------------------------------------------------
\subsection{Empirical and conceptual synthesis}

From an empirical viewpoint,
CET$\Omega$ recovers $\Lambda$CDM, GR and QFT
as limiting cases,
while offering quantitative predictions
for next-generation detectors:
LISA, Einstein Telescope, LiteBIRD, Euclid,
and Hyper-Kamiokande.
Its theoretical coherence (causal, spectral and gauge)
reaches completeness under the
\emph{Principle of Minimal Causality} (PCM),
which states that the universe minimizes its
total causal complexity under
the constraint of positive information flow.
This principle replaces the notion of
a “fundamental substance” by a
self-consistent network of causal links
that collectively generate spacetime and matter.

\begin{figure}[t]
\centering
\includegraphics[width=\linewidth]{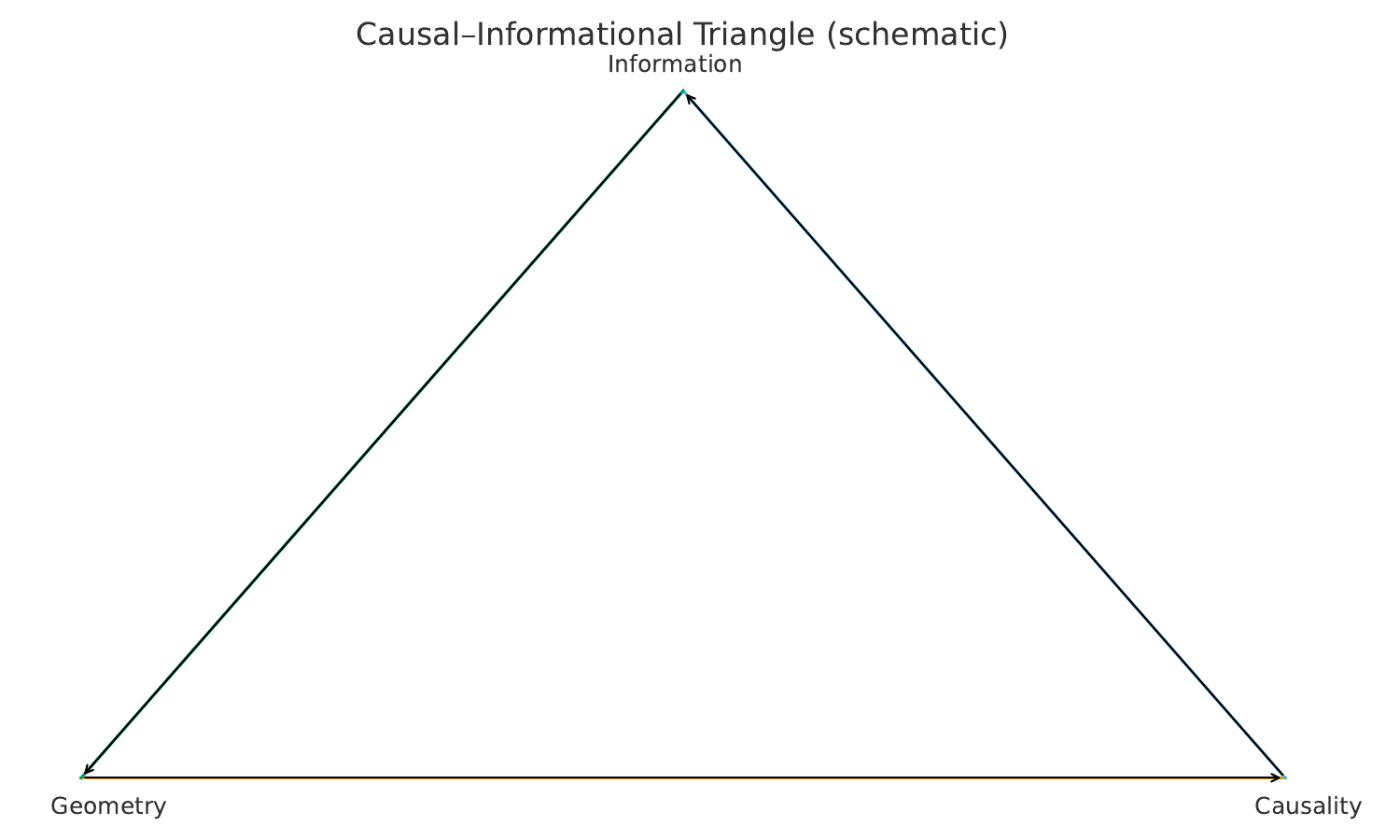}
\caption{\label{fig:tci}
Schematic representation of the
\emph{Causal–Informational Triangle (TCI)}.
Each vertex—geometry, causality and information—
mutually determines the others.
CET$\Omega$ achieves their closure
under the Principle of Minimal Causality (PCM).}
\end{figure}

%-----------------------------------------------------------
\subsection{Perspectives}

Future work will focus on three main fronts:
\begin{enumerate}
\item \textbf{Empirical validation:} confrontation of the
CET$\Omega$ parameter space with combined
CMB, LSS and gravitational-wave datasets
via the public Bayesian pipeline.
\item \textbf{Quantum information geometry:}
extension of the causal metric to the
Hilbert space of informational observables,
aiming to establish a covariant entropy–geometry duality.
\item \textbf{Foundations and philosophy:}
analysis of the ontological implications of
the Causal–Informational Completion (TCI)
for cosmology, time, and the emergence of observers.
\end{enumerate}

In conclusion, CET$\Omega$ provides a coherent,
predictive, and falsifiable completion of gravity.
It shows that the consistency of the universe
is equivalent to the completeness of its causal information.
If confirmed, it would represent a shift
from substance-based physics
to a fully informational–causal ontology.

%-----------------------------------------------------------
\section*{Acknowledgments}

The author thanks the University of Buenos Aires
for institutional support, and acknowledges the
influence of discussions with colleagues in
gravitational and quantum–information physics
that helped refine the causal framework.
This work was developed within the
CET$\Omega$ program on
\textit{Causal–Informational Unification of Physics}.

%-----------------------------------------------------------
\section*{Author Information}

\textbf{Christian Balfagón} \\
University of Buenos Aires \\
ORCID: \texttt{0009-0003-0835-5519} \\
Email: \texttt{Lyosranch@gmail.com}

%-----------------------------------------------------------
\bibliographystyle{apsrev4-2}
\bibliography{cetomega_refs}

\end{document}